# Physics-Based Triple-Debye Dielectric Modeling and Multi-Layer ADE-FDTD Simulation of Terahertz Pulse Reflection for Breast Cancer Detection

Ali Asghar Molavi Choobini[1,2], Mehran Shahmansouri[2,*]

[1]Quantum Matter Lab, Department of Physics, College of Science, University of Tehran, Tehran 14399-55961, Iran.

[2]Department of Atomic and Molecular Physics, Faculty of Physics, Alzahra University, Tehran, Iran.

* mshmansouri@gmail.com; m.shahmansouri@alzahra.ac.ir

**Abstract:**

Terahertz (THz) imaging has emerged as a promising non-ionizing modality for breast cancer assessment owing to its intrinsic sensitivity to tissue hydration. However, existing dielectric descriptions of biological tissue, largely restricted to single- or double-Debye models, fail to capture the multi-scale relaxation dynamics governing broadband THz dispersion and absorption, thereby limiting the quantitative interpretation of reflected pulse signatures.Here, a physics-based triple-Debye dielectric framework is developed to quantitatively predict broadband THz-tissue interactions. The proposed model explicitly incorporates three physically distinct relaxation processes associated with free-water rotational dynamics ($\tau_1$), bound-water relaxation ($\tau_2$), and ultrafast interfacial/macromolecular polarization ($\tau_3$). Model parameters are obtained by nonlinear least-squares fitting to experimentally measured refractive-index data digitized from published THz time-domain spectroscopy measurements of *ex vivo* human breast tissue. Compared with conventional Debye formulations, the proposed model substantially improves the fitting accuracy, reducing the root-mean-square error from 0.7773 (single Debye) and 0.2976 (double Debye) to only 0.0199 for the triple-Debye model. Full-wave FDTD-ADE simulations of realistic multilayer breast structures further demonstrate that reflected THz pulses encode tissue hydration through reproducible temporal signatures, including increased reflection amplitude, delayed pulse arrival, and enhanced waveform broadening in malignant tissue. Polarization- and angle-resolved Fresnel analysis further indicates that hydration-dependent pseudo-Brewster minima provide an additional contrast mechanism by selectively suppressing reflections from low-hydration normal and adipose tissues. Furthermore, the proposed framework accurately predicts frequency-dependent attenuation, penetration depth, reflection characteristics, and system-level imaging performance, thereby establishing a physically consistent and quantitatively validated computational framework for future experimental development of broadband THz reflection imaging for label-free breast tissue assessment.

## 1. Introduction

Terahertz (THz) waves have garnered considerable attention in the last decades owing to their distinctive interaction mechanisms with biological tissues and their non-ionizing characteristics [1- 6]. These traits make THz radiation closely resembles useful for biomedical diagnostics, such as imaging, tissue characterization, and finding cancer in its early stages. In the THz frequency range, biological tissues show a lot of dispersion and absorption. These phenomena are mostly caused by the relaxation dynamics of bound and free water molecules, as well as protein and lipid components. On the other hand, breast cancer is still one of the most common types of cancer that kills people around the world. Finding out about a disease early on is closely resembles significant for patients' chances of survival and prognosis. Realistic breast tissue has a structure with many layers, including skin, fat, fibroglandular tissue, and maybe even cancerous cells [7-12]. The dielectric properties of each layer are different. To accurately model the reflection of THz pulses off these types of stratified media, we require a multilayer electromagnetic framework that considers multiple internal reflections, impedance mismatches, and dispersive losses [13- 15]. The numerical electromagnetic simulation is crucial for understanding THz wave-tissue interactions, providing an accurate model, and optimizing diagnostic setups. The finite-difference time-domain (FDTD) method is one of the best numerical methods for simulating broadband, ultrashort THz pulse propagation in dispersive, and lossy media. Because FDTD works in the time domain, it is especially useful at using Debye-type dispersion models and looking at transient reflection signals [16- 19]. These signals are the basis for many THz time-domain spectroscopy (THz-TDS) and imaging systems.

Recent studies in THz cancer detection have shown that metamaterial absorbers and metasurface-based biosensors offer remarkable field confinement and sensitivity to even small changes in the dielectric properties of biological tissues. Semiconductor and plasmonic designs, tunable absorbers using indium antimonide and resonators on gallium arsenide substrates, have reached near-perfect absorption rates and extraordinarily high refractive index sensitivities, making it possible to distinguish breast and skin cancer tissues numerically through shifts in resonance frequency and differences in absorption [20, 21]. Multiband and compact designs that work over a wide range of THz frequencies have taken this even further by using multiple resonant modes with high quality factors and figures of merit. This shows that they work well for biological sensing applications [22, 23]. Graphene-based tunable metasurfaces and multilayer graphene absorbers have the advantage of being able to be controlled electrically by changing the chemical potential. This makes it easier to tell the difference between healthy and cancerous breast tissues [24, 25]. Cavity-enhanced metamaterials, defective photonic crystals, photonic crystal fibers, and microstrip or array antennas are other promising methods. All these studies have shown that factors including resonance frequency shifts, return loss, and reflection coefficients are closely resembles sensitive to the dielectric properties and dispersion of tissues, especially in multi-layer phantoms that imitate breast or skin structures [26- 28]. Dielectric spectroscopy and imaging experiments in the microwave and THz frequencies have repeatedly demonstrated distinct variations in permittivity between healthy and malignant breast tissues, both in patient-derived and ex vivo samples [29, 30]. In all of them, a critical requirement for any THz-based diagnostic strategy is to accurately simulate how tissue behaves as a dielectric across frequencies. In THz spectroscopy, metamaterial sensors, and antenna-based detection systems, single- and double-Debye models have been frequently employed to represent the dispersive permittivity of breast and skin tissues. These models often match up well with real-world data and have helped in tissue classification [31, 32,]. Nonetheless, numerous studies have shown that traditional Debye models inadequately represent the response of heterogeneous tissues, where fluctuations in water content, lipid composition, and structural complexity are significant factors. Because of this problem, better fitting methods and hybrid dielectric models have been created [33, 34]. Time-domain simulations based on the ADE-FDTD formulation have significantly improved the understanding of THz pulse propagation and reflection in dispersive multilayer breast tissue models. These simulations show various temporal signatures and reflection patterns that can tell normal tissues from malignant ones in both the THz and microwave ranges [35]. These simulations also indicate how elements like tissue layers, the chest wall, the precision of dispersion models, and the positioning of antennas can have a large effect on how well detection works. Progress has sped up even further thanks to advanced anatomical phantoms, reduced-order inversion techniques, and multimodal or data-driven frameworks. All these things have made breast cancer imaging more realistic and better at categorizing [36, 37].

Although single- and double-Debye models are often employed to predict the dispersive permittivity of biological tissues in the THz range, they generally fail to capture the complex multi-scale dielectric relaxation behavior of heterogeneous breast tissue. The proposed model incorporates three physically distinct dielectric relaxation processes associated with free-water rotational dynamics, bound-water relaxation, and ultrafast interfacial/macromolecular polarization. The corresponding relaxation times are determined through nonlinear optimization of the experimental dielectric spectra rather than being prescribed a priori. Experimental THz dielectric spectroscopy of ex vivo breast samples consistently demonstrates that lower-order Debye models produce larger residual errors, particularly in the intermediate-frequency region where multiple relaxation mechanisms contribute simultaneously. Building on prior THz diagnostic frameworks, this study introduces a Triple-Debye composite dielectric model integrated into a multilayer FDTD–ADE simulation for time-domain THz pulse reflection from realistic breast tissue structures. The proposed Triple-Debye formulation is strongly supported by experimental THz dielectric spectroscopy, which consistently reveals multi-step dispersion behavior and broad absorption features in breast tissues that cannot be reproduced by lower-order Debye models. By integrating the proposed triple-Debye formulation into a multilayer FDTD model, explicitly accounting for breast, adipose, fibroglandular, and tumor layers with distinct dispersive parameters, the present work provides a more physiologically realistic representation of THz pulse propagation, multiple internal reflections, and impedance mismatches in stratified breast media. In particular, adding the additional slow relaxation pole associated with hydration shells and macromolecular interactions leads to increase of systematic of both the real and imaginary parts of permittivity, especially in the THz ranges that are critical for pulse-based reflection analysis. This advanced modeling framework provides enhanced predictions of temporal waveforms and reflection coefficients in comparison to lower-order Debye models, facilitating the precise differentiation of subtle dielectric contrasts between healthy and malignant tissues. The suggested framework sets up a physically consistent and quantitatively accurate model for THz pulse reflection. This makes it more sensitive to early-stage breast cancer and provides a strong base for improving and understanding reflection-based THz-TDS diagnostic systems. The paper is organized as follows: In section II, the theory of the triple-Debye composite dielectric model is presented. Results and discussion is considered in section III. Conclusions are drawn in section IV.

# 2. Theoretical Model

To simulate realistic reflection-mode THz imaging scenarios, including ex vivo breast tissue characterization, intra-operative margin assessment, and laboratory phantom measurements, the multi-layer stratified configuration depicted in Fig. (1) is considered. The structure consists of four distinct regions: incident air (Region 1), a breast tissue layer of thickness $d_1$ (Region 2), a quartz glass substrate of thickness $d_2$ (Region 3), and a terminating air region (Region 4). An obliquely incident THz pulse impinges on the tissue interface at an angle $\theta_1$, undergoing multiple refractions and reflections governed by the frequency-dependent complex permittivity of each layer. The resulting multi-interface interference and time-delayed reflections produce a characteristic reflected time-domain waveform that is highly sensitive to tissue dielectric properties, layer thicknesses, and angular incidence conditions. Its implementation within the FDTD framework using the auxiliary differential equation (ADE) approach, the parameter extraction strategy for biological tissue, and an analytical multi-layer reflection model employed for numerical validation are presented in the following sections.

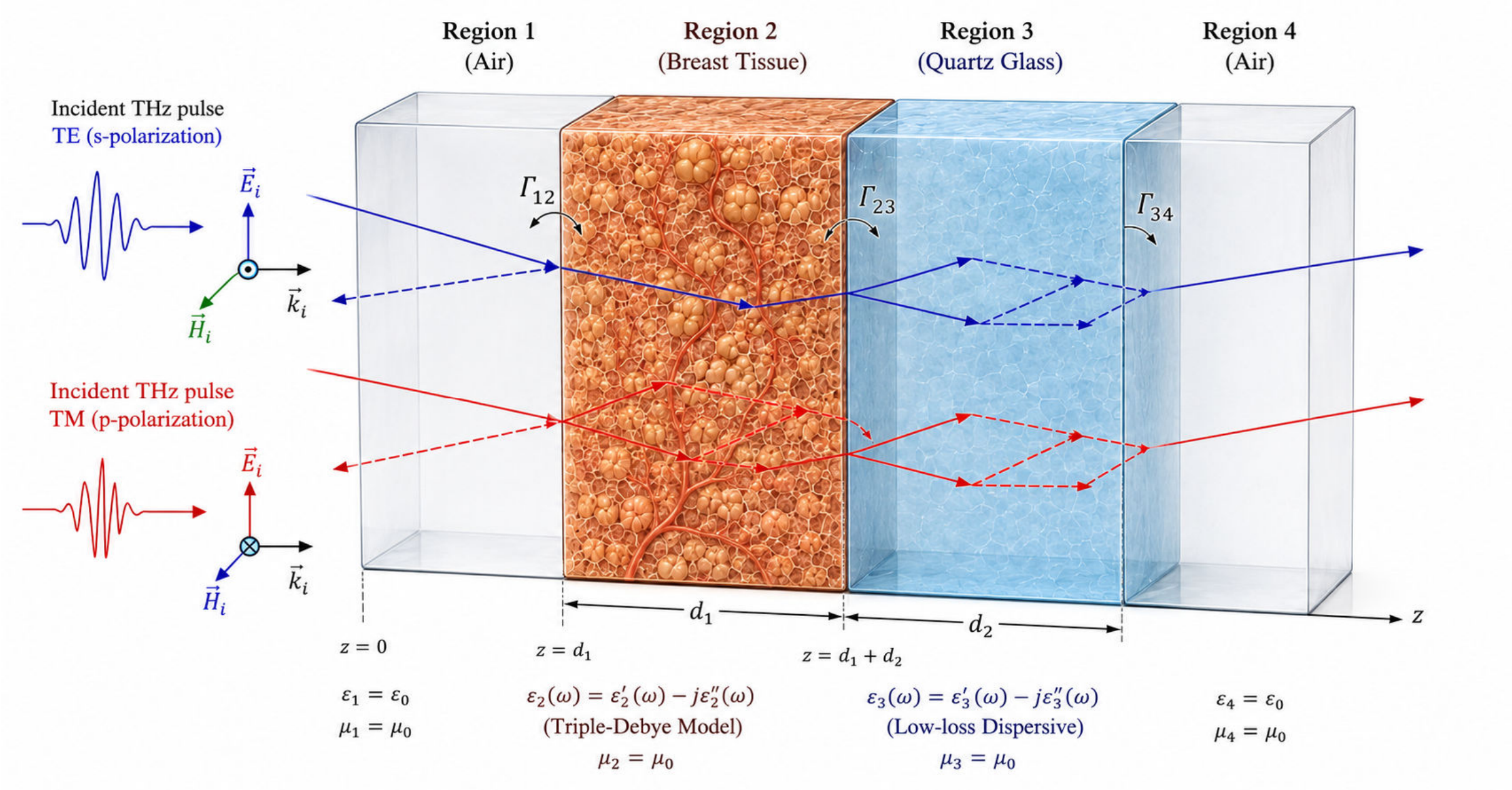

Figure 1: Schematic of oblique incidence THz pulse reflection in a multi-layer configuration.

## 2.1 Triple-Debye Complex Permittivity Model

A triple-Debye relaxation model is adopted as the minimum physically accurate representation capable of simultaneously accounting for fast free-water relaxation, intermediate bound-water dynamics, and slower macromolecular polarization contributions. Therefore, the general expression of the Triple-Debye model can be written as:

$$\hat{\epsilon}(\omega) = \epsilon_\infty + \sum_{k=1}^{3} \frac{\Delta\epsilon_k}{1+i\omega\tau_k}, \tag{1}$$

where $\hat{\epsilon}(\omega) = \epsilon'(\omega) + i\epsilon''(\omega)$ is the complex permittivity. The parameter $\epsilon_\infty$ signifies the permittivity at elevated frequencies, encompassing instantaneous electronic polarization and high-frequency vibrational contributions that respond almost quickly to the applied electromagnetic field. The quantity of $\Delta\epsilon_k$ represent the dielectric strengths linked to the $k$th relaxation mechanism and measure how much each polarization process adds to the total static permittivity. The parameters $\tau_k$ are the relaxation times that correspond to each polarization mechanism. They indicate how long it takes for each mechanism to respond to changes in the electric field. The angular frequency is $\omega = 2\pi\nu$, where $\nu$ is the linear frequency. Equation (1) explicitly indicates that each polarization mechanism reacts differently across the THz frequency band, which causes a cumulative dispersive behavior. To make things clearer, the expanded version of the TripleDebye model is:

$$\hat{\epsilon}(\omega) = \epsilon_\infty + \frac{\Delta\epsilon_1}{1+i\omega\tau_1} + \frac{\Delta\epsilon_2}{1+i\omega\tau_2} + \frac{\Delta\epsilon_3}{1+i\omega\tau_3}, \tag{2}$$

which explicitly separates the contributions of the three relaxation poles. At greater THz frequencies, the fastest relaxation process, which is usually linked to free water molecules, is the most important. At lower frequencies, the slower relaxation processes are the most significant.

To find the static permittivity, which is the same as the dielectric response at zero frequency, limit of the dielectric response, the set $\omega \to 0$ is assumed in Eq. (1). In this limit, the frequency-dependent terms vanish and the static permittivity is given by:

$$\epsilon_S = \epsilon_\infty + \Delta\epsilon_1 + \Delta\epsilon_2 + \Delta\epsilon_3. \quad (3)$$

This relation indicates that the static permittivity arises from the cumulative contribution of all polarization mechanisms. The complex permittivity is related to the complex refractive index $\hat{n}(\omega)$ through the constitutive relation as follows:

$$\hat{\epsilon}(\omega) = \hat{n}(\omega)^2, \quad (4)$$

which follows directly from Maxwell's equations for linear, isotropic, and non-magnetic media, links the complex permittivity $\hat{\epsilon}(\omega)$ to the complex refractive index $\hat{n}(\omega)$. The complex refractive index is expressed as:

$$\hat{n}(\omega) = n(\omega) + i\kappa(\omega), \quad (5)$$

where $n(\omega)$ is the refractive index governing phase propagation, and $\kappa(\omega)$ is the extinction coefficient describing electromagnetic energy dissipation due to absorption within the medium. The extinction coefficient is related to the absorption coefficient $\alpha(\nu)$ by:

$$\kappa(\nu) = \frac{c\,\alpha(\nu)}{2\pi\nu}, \quad (6)$$

where $c$ is the speed of light in vacuum, and $\nu$ denotes the frequency. This expression directly links the imaginary part of the refractive index to the exponential attenuation of electromagnetic wave amplitude as it propagates through the medium. Substituting this definition into the constitutive relation and separating the real and imaginary components yields:

$$\epsilon'(\nu) = n(\nu)^2 - \kappa(\nu)^2, \quad (7)$$

$$\epsilon''(\nu) = 2n(\nu)\kappa(\nu). \quad (8)$$

The real part $\epsilon'$ governs wave propagation and phase velocity, whereas the imaginary part $\epsilon''$ quantifies dielectric losses and directly determines THz absorption in biological tissue. The Triple-Debye model contains seven independent parameters as follows:

$$\{\epsilon_\infty, \Delta\epsilon_1, \Delta\epsilon_2, \Delta\epsilon_3, \tau_1, \tau_2, \tau_3\},$$

which are determined by fitting the model to experimentally measured permittivity spectra. This fitting procedure is carried out by minimizing the objective function

$$S = \sum_{j=1}^{N} \left[ \left(\epsilon'_{\text{model}}(\nu_j) - \epsilon'_{\text{meas}}(\nu_j)\right)^2 + \left(\epsilon''_{\text{model}}(\nu_j) - \epsilon''_{\text{meas}}(\nu_j)\right)^2 \right], \quad (9)$$

where $\nu_j$ denotes the $j$th frequency sampling point and $N$ is the total number of frequency points. This cost function ensures that both the dispersive and absorptive parts of the dielectric response are captured at the same time. The model depends on the relaxation durations τ k in a nonlinear way, hence nonlinear optimization methods like the Levenberg–Marquardt algorithm are used.

## 2.2 Maxwell–ADE Formulation for Triple-Debye Media

To incorporate the dispersive Triple-Debye permittivity into time-domain simulations, Maxwell's curl equations are coupled with auxiliary differential equations describing the temporal evolution of polarization. Unlike an empirical numerical treatment, the auxiliary differential equation (ADE) formulation employed in this work is derived directly from the frequency-domain Debye constitutive relation. For a single Debye relaxation process, the constitutive relation is expressed as

$$\hat{\epsilon}(\omega) - \epsilon_\infty = \frac{\Delta\epsilon}{1+i\omega\tau}. \quad (10)$$

The polarization density in the frequency domain is related to the electric field through:

$$\mathbf{P}(\omega) = \epsilon_0[\hat{\epsilon}(\omega) - \epsilon_\infty]\mathbf{E}(\omega). \quad (11)$$

Substituting Eq. (10) into Eq. (11) yields

$$(1 + i\omega\tau)\, \mathbf{P}(\omega) = \epsilon_0 \Delta\epsilon\, \mathbf{E}(\omega). \tag{12}$$

Applying the inverse Fourier transform and using the correspondence $i\omega \leftrightarrow \frac{\partial}{\partial t}$, the above frequency-domain relation is transformed into the first-order differential equation governing the polarization density,

$$\tau \frac{\partial \mathbf{P}}{\partial t} + \mathbf{P} = \epsilon_0 \Delta\epsilon\, \mathbf{E}. \tag{13}$$

For the Triple-Debye model, the same derivation is applied independently to each relaxation process, resulting in three auxiliary polarization equations corresponding to the three relaxation mechanisms. These auxiliary equations constitute the time-domain representation of the dispersive dielectric response and are subsequently coupled with Maxwell's curl equations within the ADE-FDTD framework. Consequently, the numerical formulation preserves the causal frequency-dependent behavior of biological tissues while remaining fully consistent with the underlying Triple-Debye constitutive model. In three dimensions, Maxwell's curl equations are written as:

$$\nabla \times \mathbf{H} = \epsilon_0 \epsilon_\infty \frac{\partial \mathbf{E}}{\partial t} + \sum_{k=1}^{3} \frac{\partial \mathbf{P}_k}{\partial t}, \tag{14}$$

$$\nabla \times \mathbf{E} = -\mu_0 \frac{\partial \mathbf{H}}{\partial t}. \tag{15}$$

Here, $\mathbf{P}_k$ represents the polarization density associated with the $k$th Debye relaxation process. The first term on the right-hand side accounts for instantaneous polarization, while the summation term captures delayed polarization effects. Each polarization component satisfies the auxiliary differential equation:

$$\tau_k \frac{\partial \mathbf{P}_k}{\partial t} + \mathbf{P}_k = \epsilon_0 \Delta\epsilon_k\, \mathbf{E}, \tag{16}$$

which enforces causality and ensures that the time-domain formulation properly reproduces the frequency-domain Debye response. The polarization equations are discretized in time using a leapfrog approach within the FDTD framework. The update equation for the $k$th polarization component is given by:

$$\mathbf{P}_k^{n+1/2} = C_{a,k}\, \mathbf{P}_k^{n-1/2} + C_{b,k}\, \mathbf{E}^{n}, \tag{17}$$

where the update coefficients are defined as:

$$C_{a,k} = \frac{2\tau_k - \Delta t}{2\tau_k + \Delta t}, C_{b,k} = \frac{2\epsilon_0 \Delta\epsilon_k \Delta t}{2\tau_k + \Delta t}. \tag{18}$$

Here, $\Delta t$ is the FDTD time step. The electric field update equation is written as

$$\mathbf{E}^{n+1} = \mathbf{E}^{n} + \frac{\Delta t}{\epsilon_0 \epsilon_\infty} \left[ (\nabla \times \mathbf{H})^{n+1/2} - \sum_{k=1}^{3} \frac{\mathbf{P}_k^{n+1/2} - \mathbf{P}_k^{n-1/2}}{\Delta t} \right], \tag{19}$$

which ensures that broad-band THz pulses move through dispersive biological mediums in a steady and precise way.

# 3 Results & Discussion

This section presents the outcomes of multi-layer FDTD simulations utilizing a triple-Debye composite dielectric framework to precisely model broadband THz interactions with heterogeneous breast tissue [38]. This method overcomes the limitations associated with conventional double-Debye models by explicitly incorporating three distinct relaxation poles corresponding to fast free-water rotational dynamics, intermediate bound-water processes, and ultrafast interfacial/macromolecular polarization mechanisms. The proposed model provides a physically substantiated description of hydration-induced variations in permittivity and absorption, enabling accurate differentiation among tumour, normal fibroglandular, and adipose tissues [39, 40].

The numerical simulations were performed within the FDTD framework using the auxiliary differential equation (ADE) method to rigorously model dispersive polarization dynamics in the time domain [41- 43]. The stratified computational domain consists of four regions: incident air (Region 1, $\varepsilon = 1$), a breast tissue layer of variable thickness $d_1$ (Region 2, governed by the optimized triple-Debye dielectric parameters), a quartz substrate with a fixed thickness of $d_2 = 1$ mm (Region 3, $\varepsilon = 3.8$, weakly dispersive), and a terminating air region (Region 4). The computational domain is surrounded on all sides by perfectly matched layer (PML) absorbing boundaries to suppress artificial reflections. The structure is excited by an obliquely incident Gaussian-modulated broadband THz plane wave with a center frequency near 1 THz and a spectral bandwidth spanning 0.1–10 THz. The angle of incidence, $\theta_1$, is varied from $0°$ to $60°$ depending on the simulation scenario, and both transverse electric (TE) and transverse magnetic (TM) polarizations are considered to investigate polarization-dependent Fresnel reflection characteristics and Pseudo-Brewster-angle phenomena. The spatial discretization was chosen as $\Delta x = \Delta y = 1.5\ \mu$m, corresponding to $\lambda_{\min}/20$ at the highest simulated frequency of 10 THz, where $\lambda_{\min} = 30\ \mu$m. The temporal step was selected according to the Courant stability criterion, resulting in $\Delta t = 0.002$ ps. Reflection and transmission monitors were positioned immediately before and after the tissue interface to record the transient electric fields for subsequent frequency-domain analysis. The principal numerical parameters used throughout the simulations are summarized in Table 1. The experimental refractive-index spectra reported by Fitzgerald *et al.* were digitized using WebPlotDigitizer, and the extracted data were subsequently employed for nonlinear least-squares optimization of the proposed triple-Debye dielectric model. The dielectric parameters were not selected empirically. Instead, they were determined through nonlinear least-squares optimization by simultaneously estimating $\varepsilon_\infty$, $\Delta\varepsilon_i$, and $\tau_i$ so as to minimize the least-squares difference between the measured and simulated refractive-index spectra over the entire measurement bandwidth. The optimized dielectric parameters employed throughout the subsequent FDTD simulations are summarized in Table 4. The quality of the nonlinear fitting was evaluated using both the root-mean-square error (RMSE) and the coefficient of determination ($R^2$), confirming the superior agreement of the proposed triple-Debye model with the experimental spectra compared with conventional single- and double-Debye formulations. Compared with normal breast tissue, the optimized dielectric parameters for tumour tissue exhibit systematically larger dielectric strengths and shorter characteristic relaxation times, consistent with the increased water content of malignant tissue. In contrast, adipose tissue is characterized by substantially reduced dielectric strengths, reflecting its relatively low hydration level.

Table 1: Numerical parameters employed in the two-dimensional FDTD simulations.

| Parameter | Value |
|---|---|
| Computational domain | $6 \times 4$ mm$^2$ |
| Spatial discretization | $\Delta x = \Delta y = 1.5\ \mu$m |
| Temporal step | $\Delta t = 0.002$ ps |
| Boundary condition | Perfectly Matched Layer (PML) |
| PML thickness | 20 cells |
| Excitation source | Gaussian-modulated plane wave |
| Frequency range | 0.1–10 THz |
| Incidence angle | $45°$ (unless otherwise specified) |
| Reflection monitor | 0.5 mm before the tissue interface |
| Transmission monitor | 0.5 mm after the tissue interface |
| Tissue thickness | 1.2 mm |

Figure 2 shows a series of snapshots of the out-of-plane electric field component ($E_z$) at different times. These snapshots show how incident radiation, interfacial reflections, and dispersive absorption interact with each other in the heterogeneous medium. At t = 5.00 ps, the modulated THz pulse hits the air-tissue interface and appears as a tightly localized wavefront with a high amplitude (peak normalized $|E| \approx 0.04$). The ring-like pattern is caused by the cylindrical symmetry of the point-like excitation source in the 2D computational domain. The leading-edge field concentrations show that refraction into the higher-index tissue layer is starting. At t = 10.00 ps, the pulse has partially entered the tissue, and it is distinct that dispersive broadening is starting to happen. The wavefront shows strong trailing oscillations and a lower peak amplitude. This is due to the frequency-dependent relaxation processes that the triple-Debye model captures, especially the slow free-water and intermediate bound-water terms that are responsible for energy dissipation in the lower THz band. At the same time, a reflected part moves backward into free space, and internal reflections at the tissue-substrate interface start to show up as faint secondary wavefronts. At t = 15.00 ps, the big pulse distortion is clear deeper into the tissue layer. The main wavefront has spread out a lot, and the energy is spread out over a longer time period because the phase velocities are different across the pulse spectrum. Multiple internal reflections create secondary pulses that interfere with each other, resulting in complex interference fringes that look like bands of alternating high and low fields. These characteristics highlight the time-domain waveform's sensitivity to minor changes in dielectric properties. Malignant tissue regions, which have greater water content and stronger relaxation strengths, would show faster attenuation and different reflection timing than healthy tissue that is mostly made up of fat. At t = 20.00 ps, the pulse has spread out and been absorbed a lot. The leftover energy shows up as low-amplitude, long, oscillating tails throughout the tissue volume. The constant interference patterns show how the quartz substrate traps and re-emits radiation, which leads to late-time echoes that are diagnostic signatures in reflection-mode measurements.

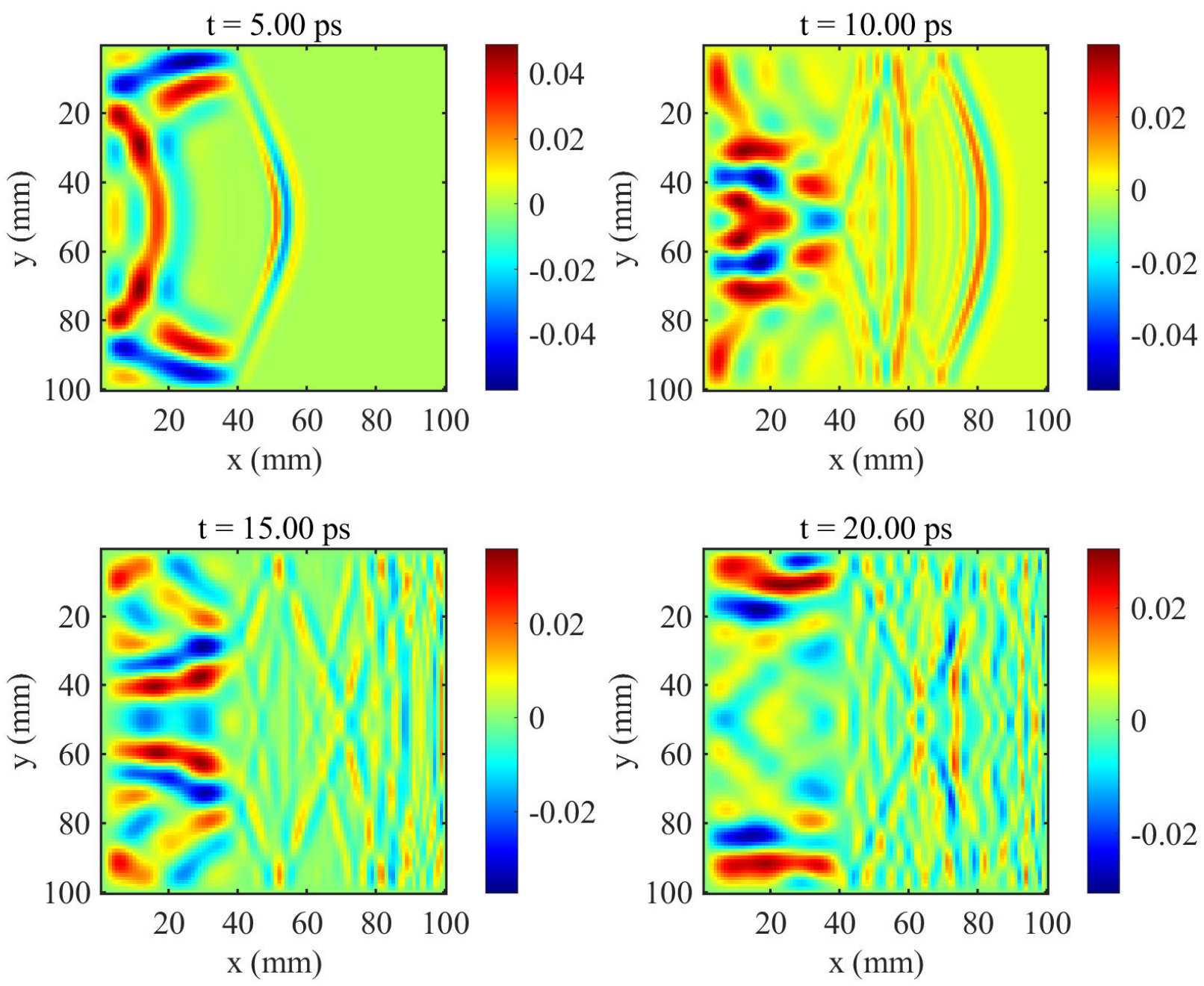

Figure 2: Various snapshots of time evolution of the electric field distribution during propagation of a THz pulse through a multi-layer breast tissue phantom in 2D FDTD simulation.

The nonlinear least-squares fitting was performed using broadband THz time-domain spectroscopy (THz-TDS) measurements of ex vivo human breast tumour tissue to evaluate the accuracy of the proposed Triple-Debye dielectric model. Figure 3 compares the experimental refractive-index data with the fitted responses obtained from the Single, Double, and Triple-Debye models over the frequency range of 0.1–2 THz, together with the corresponding time-domain reflected THz waveforms shown in the inset. The experimental refractive-index data were obtained by carefully digitizing the measurements reported by Fitzgerald et al. [44] in their pioneering study on THz reflection modelling of breast tissue using Debye theory and FDTD simulations. The extracted dataset exhibits the characteristic monotonic decrease in refractive index from approximately 2.33 at 0.1 THz to about 1.65 at 2 THz, which is consistent with the dielectric dispersion expected for malignant breast tissue. A total of approximately 500 digitized data points were extracted from the published spectra using WebPlotDigitizer and were directly used in the fitting procedure without any interpolation or downsampling in order to preserve the original experimental characteristics. The dielectric parameters of the Single-, Double-, and Triple-Debye models were estimated via nonlinear least-squares optimization using the Levenberg–Marquardt algorithm by minimizing the residual error between the experimental and modeled refractive-index spectra. Equal weighting was assigned to all frequency samples due to the absence of reported measurement uncertainties in the original study. Since only the refractive-index dispersion was available from the published data, the optimization was performed exclusively on the real refractive-index response. The same overall frequency-dependent behaviour has also been independently reported in subsequent experimental THz-TDS studies on breast tissue [45- 48], supporting the general consistency of the dielectric response across different measurement systems and tissue samples. The Single-Debye model shows the largest deviation from the experimental data throughout the investigated frequency range. Because only one dielectric relaxation process is included, the model is unable to fully capture both the rapid low-frequency dispersion and the gradual high-frequency decay observed experimentally. Consequently, the predicted refractive index is systematically lower than the experimental data, particularly in the intermediate-frequency region where multiple polarization mechanisms coexist. The Double-Debye model significantly improves the agreement by introducing a second relaxation process associated with additional polarization mechanisms, including bound-water effects. Nevertheless, noticeable discrepancies remain over the entire spectrum, and the model still predicts a steeper reduction in refractive index than observed experimentally, indicating that two relaxation mechanisms are insufficient to fully describe the complex dielectric behaviour of malignant breast tissue. In contrast, the Triple-Debye model provides excellent agreement with the experimental measurements across the entire frequency interval, as also evidenced in Fig. 3. The fitted curve closely follows the measured data points and accurately reproduces both the initial rapid decrease below approximately 0.5 THz and the gradual reduction at greater frequencies. The quantitative fitting results further confirm this improvement, with RMSE values of 0.7773, 0.2976, and 0.0199 for the Single-, Double-, and Triple-Debye models, respectively. Compared with the conventional Single-Debye formulation, the proposed Triple-Debye model reduces the fitting error by approximately $97.4\%$, while achieving a $93.3\%$ improvement relative to the Double-Debye model. These results demonstrate that the additional relaxation process incorporated in the Triple-Debye formulation is essential for accurately describing the broadband dielectric response of breast tumour tissue. The inset further illustrates the corresponding normalized reflected THz pulses obtained using the fitted dielectric models. Increasing the number of relaxation processes produces progressively larger temporal dispersion, increased pulse broadening, reduced peak amplitude, and a measurable propagation delay. The Triple-Debye waveform therefore exhibits the broadest temporal profile and the strongest attenuation, reflecting the enhanced frequency-dependent phase delay and dielectric losses predicted by the fitted material parameters. This close agreement between the frequency-domain fitting and the resulting time-domain response indicates the self-consistency of the proposed dielectric model and provides greater confidence in its application to full-wave electromagnetic simulations of multilayer breast tissue structures. To further quantify the fitting performance, the optimized dielectric parameters obtained from the nonlinear least-squares procedure are summarized in Table 2. All optimized parameters remain within physically meaningful bounds reported for biological tissues and are consistent with the dielectric relaxation mechanisms of breast tissue constituents.

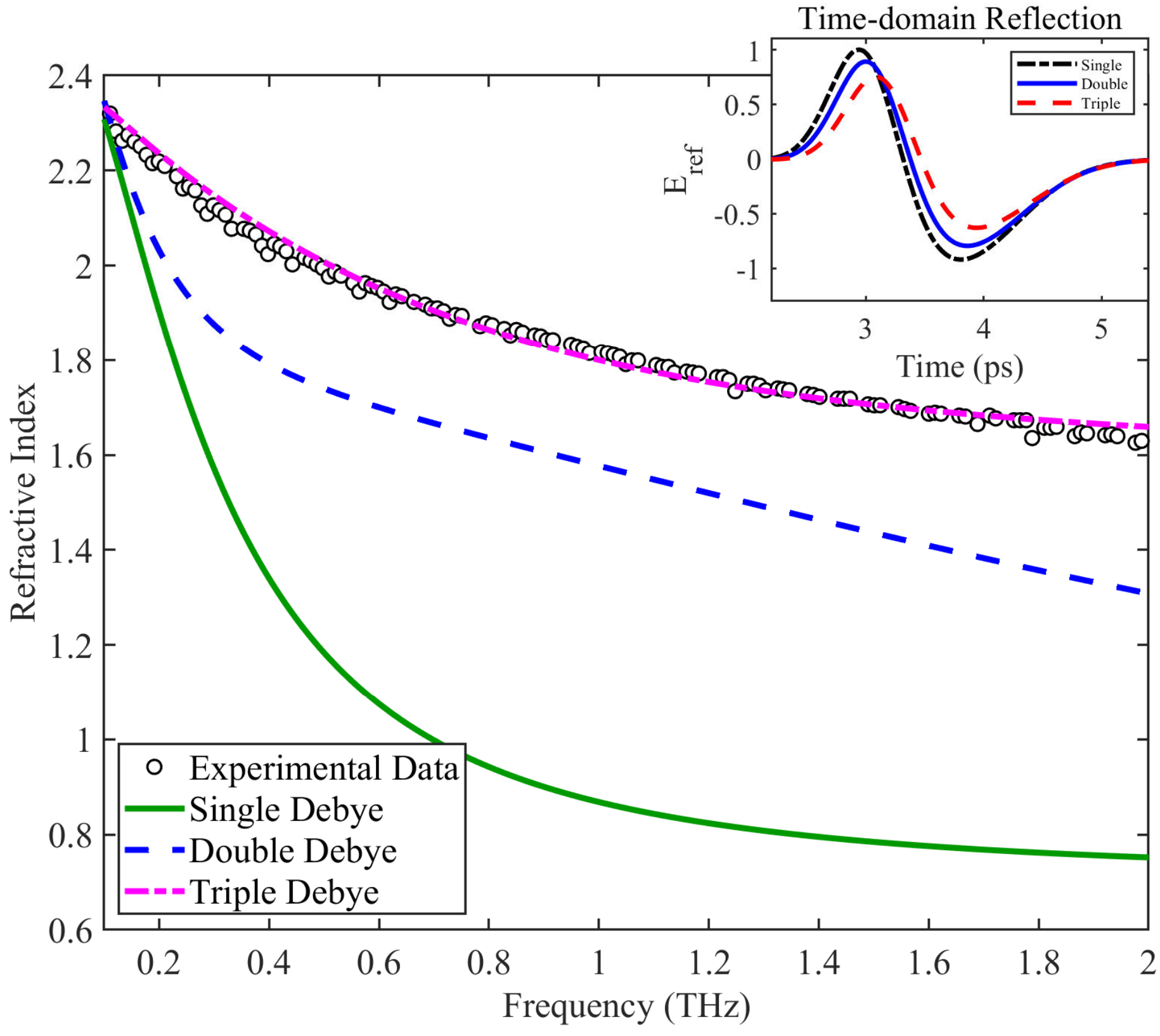

Figure 3: Experimental refractive index of ex vivo human breast tumour tissue digitized [44] together with the fitted three Debye models over the 0.1–2 THz range. The inset presents the corresponding normalized time-domain reflected THz waveforms predicted by the three dielectric models.

Table 2: Fitted triple-Debye dielectric parameters obtained by nonlinear least-squares fitting of digitized experimental THz refractive-index data for normal and tumour breast tissues.

| **Parameter** | **Normal Tissue** | **Tumour Tissue** | **Unit** | **Physical Interpretation** | **Search Range** |
|---|---|---|---|---|---|
| $\varepsilon_\infty$ | 2.5 | 3.0 | – | High-frequency permittivity | 1–10 |
| $\Delta\varepsilon_1$ | 48.8 | 55.1 | – | Free-water relaxation strength | 20–80 |
| $\Delta\varepsilon_2$ | 14.86 | 20.04 | – | Bound-water relaxation strength | 5–30 |
| $\Delta\varepsilon_3$ | 2.05 | 4.97 | – | Interfacial/macromolecular polarization strength | 0–10 |
| $\tau_1$ | 12.0 | 10.0 | ps | Free-water rotational relaxation time | 5–15 |
| $\tau_2$ | 0.50 | 0.60 | ps | Bound-water relaxation time | 0.1–2.0 |
| $\tau_3$ | 0.08 | 0.09 | ps | Macromolecular relaxation time | 0.01–0.20 |

Figure 4 further illustrates the electromagnetic properties derived from the optimized dielectric models. Panels (a) and (b) present the frequency-dependent refractive index and absorption coefficient of water, tumour, normal, and adipose tissues over the THz band, while panel (c) directly compares the real permittivity predicted by the Single, Double, and Triple-Debye formulations for tumour tissue. The inset in panel (c) quantifies the relative deviation of the simplified models with respect to the proposed Triple-Debye model. As shown in Figure 4(a), the refractive index decreases monotonically with increasing frequency for all tissues, which is the characteristic signature of dielectric relaxation. Tumour tissue consistently exhibits a larger refractive index than normal breast tissue throughout the investigated frequency range due to its greater water content and stronger molecular polarization. Water maintains the highest refractive index at low frequencies owing to its dominant free-water relaxation, whereas adipose tissue shows the weakest dispersion and rapidly approaches its high-frequency limit due to its low water fraction and limited dielectric polarization. The absorption characteristics shown in Figure 4(b) demonstrate the complementary behaviour. The absorption coefficient increases with frequency for every tissue as dielectric losses become more pronounced. Tumour tissue exhibits substantially stronger attenuation than normal tissue over the entire spectral range, reflecting the larger dielectric loss associated with increased hydration and interfacial polarization. Water presents high absorption at low and intermediate frequencies because of molecular rotational relaxation, whereas adipose tissue displays comparatively weaker attenuation at low frequencies but gradually approaches the tumour response at greater frequencies owing to the frequency dependence of its dielectric loss. These differences in attenuation directly contribute to the contrast observed in THz reflection measurements. The physical advantage of the Triple-Debye formulation becomes more evident in Figure 4(c). The Single-Debye model predicts an excessively rapid decrease in the real permittivity and fails to reproduce the gradual transition observed experimentally, indicating that a single relaxation mechanism cannot adequately describe the broadband dielectric response of breast tissue. The Double-Debye model significantly improves the overall behaviour by introducing a second relaxation process; however, noticeable discrepancies remain over the intermediate-frequency region, where multiple polarization mechanisms overlap. In contrast, the proposed Triple-Debye model produces a smooth multi-stage relaxation profile that accurately captures the broadband dielectric dispersion expected for heterogeneous biological tissue containing free water, bound water, and interfacial polarization processes. The inset quantitatively highlights these differences by presenting the relative deviation of the simplified models with respect to the Triple-Debye solution. The Single-Debye approximation exhibits deviations approaching approximately $60\%$ over a substantial portion of the spectrum, demonstrating that a single relaxation pole cannot reproduce the complex dielectric response of malignant breast tissue. Although the Double-Debye model considerably reduces the discrepancy, deviations of approximately $20-30\%$ remain around the principal relaxation region, where both free-water and bound-water dynamics contribute simultaneously. These results provide quantitative evidence that the additional relaxation pole introduced by the Triple-Debye model is not merely a mathematical refinement but a physically meaningful extension required to reproduce the broadband dielectric behaviour of breast tissue. Figures 3 and 4 collectively establish the validity of the proposed dielectric model. The excellent agreement with experimentally digitized refractive-index measurements, together with the physically consistent evolution of the refractive index, absorption coefficient, and permittivity spectra, demonstrates that the optimized Triple-Debye parameters accurately describe the multi-scale dielectric relaxation mechanisms governing THz wave propagation in breast tissue.

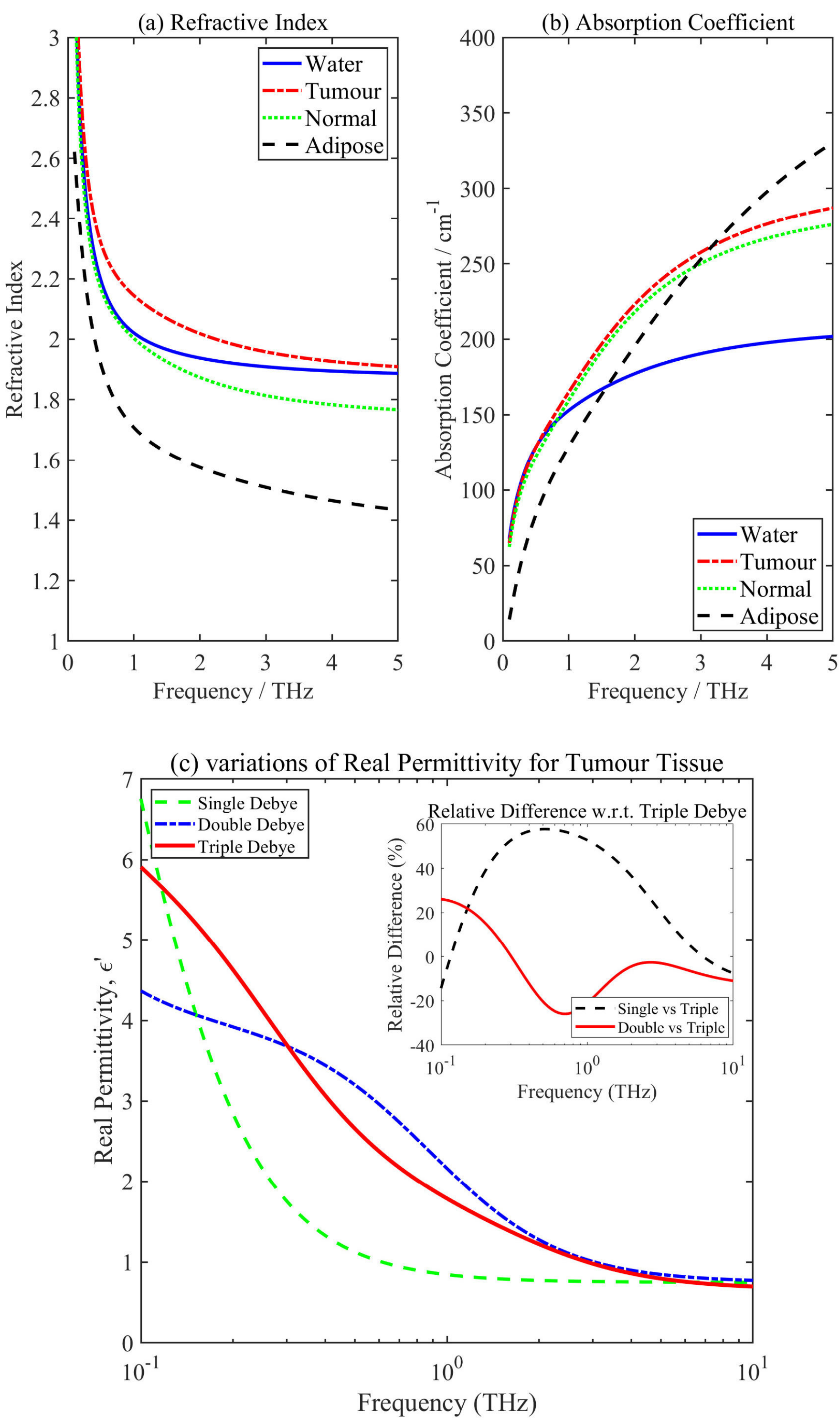

Figure 4: Frequency-dependent optical properties predicted by the proposed Debye models. (a) Refractive index of water, tumour, normal, and adipose breast tissues. (b) Corresponding absorption coefficient over the 0.1–5 THz frequency range. (c) Comparison of the real permittivity predicted by the three Debye models for tumour tissue, together with the relative deviation of the simplified models from the Triple-Debye

solution (inset).

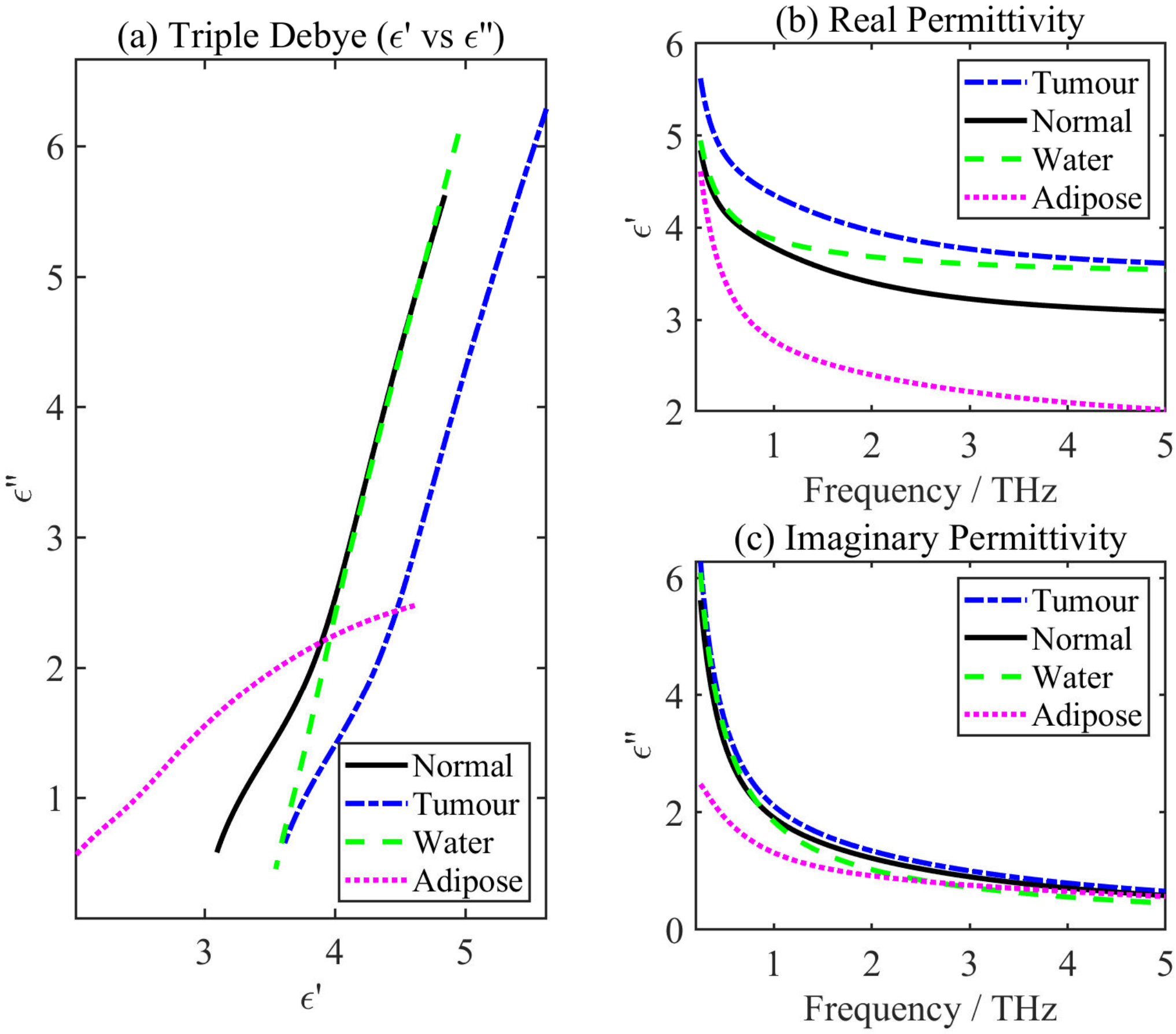

Figure 5: Complex permittivity dispersion derived from the triple-Debye model for normal and tumour breast tissue, compared to pure water and adipose tissue references.

Figure 6 shows normalized reflected electric field pulses for different types of tissue. The fitted triple-Debye parameters show distinct differences in shape. The tumor waveform has the biggest first positive peak (about 0.55 a.u.) and the lowest second negative peak (about −0.25 a.u.). The stronger Fresnel reflection at the air–tumour interface results from the greater refractive index and enhanced dielectric dispersion associated with increased water content. Normal breast tissue has a much lower primary reflection amplitude (about 0.45 a.u.) and a shallower inversion, which is what you would expect from fibroglandular regions with moderate levels of hydration. Pure water shows the strongest response, with the tallest peak and deepest trough in its waveform, coming closely resembles close to the theoretical limit for a fully hydrated medium. In contrast, adipose (fat) tissue produces the weakest and least distorted pulse—its waveform stays relatively clean and simple, with almost no significant negative dip, thanks to its low permittivity and weak frequency dispersion. These differences in waveform shape stem from impedance mismatches at the boundaries between tissues, combined with the way waves propagate differently depending on frequency as they travel through the tissue layers, and greater water content. This leads to greater phase delay at low frequencies, which in turn delays the inversion of the bipolar pulse and makes the trailing oscillations more pronounced and amplified. The distinct difference between the signatures of tumor and normal tissue, especially in peak-to-peak amplitude and zero-crossing delay, shows that the triple-Debye representation can create strong contrast. This is because it captures the multi-scale relaxation that simpler models don't. Figure 7 shows how reflected pulses change in a systematic way as the hydration fraction is changed in a tumour tissue model. This is done to directly look at the source of this contrast. As the amount of water in the system goes from 0.01 to 1.00, the main reflection amplitude rises steadily, the negative peak becomes substantially deeper, and the pulse gets wider with delayed zero-crossings. Even small increases in hydration (0.25 to 0.50) cause noticeable changes in the shape of the waveform. This shows how sensitive THz reflection is to both bound and free water contributions, which is exactly what happens when cells become cancerous. The time-domain waveforms calculated with the multi-layer FDTD framework and validated analytical model make it distinct that reflection-mode THz pulse imaging has diagnostic potential. The results of the simulation show that THz pulse reflection data can be used to measure tissue hydration and dielectric relaxation. This non-ionizing and label-free imaging mechanism provides a promising physical basis for future experimental investigations of breast tissue characterization. The improved accuracy of the proposed triple-Debye model enables more reliable prediction of time-domain waveform features, thereby providing a robust computational foundation for future THz imaging system design and experimental validation.

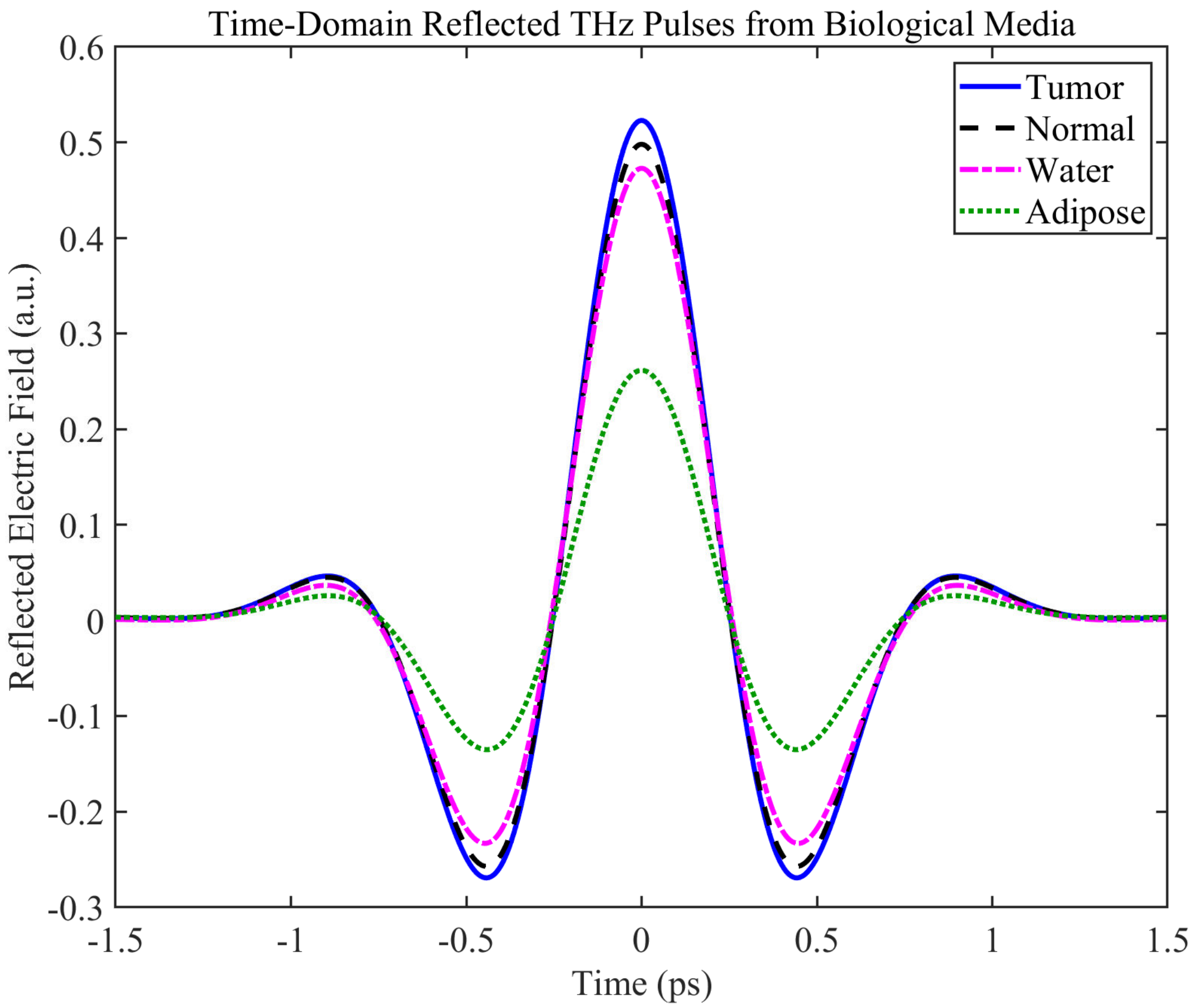

Figure 6: Simulated time-domain reflected THz pulses from multi-layer models of various tissue using the triple-Debye dispersion parameters.

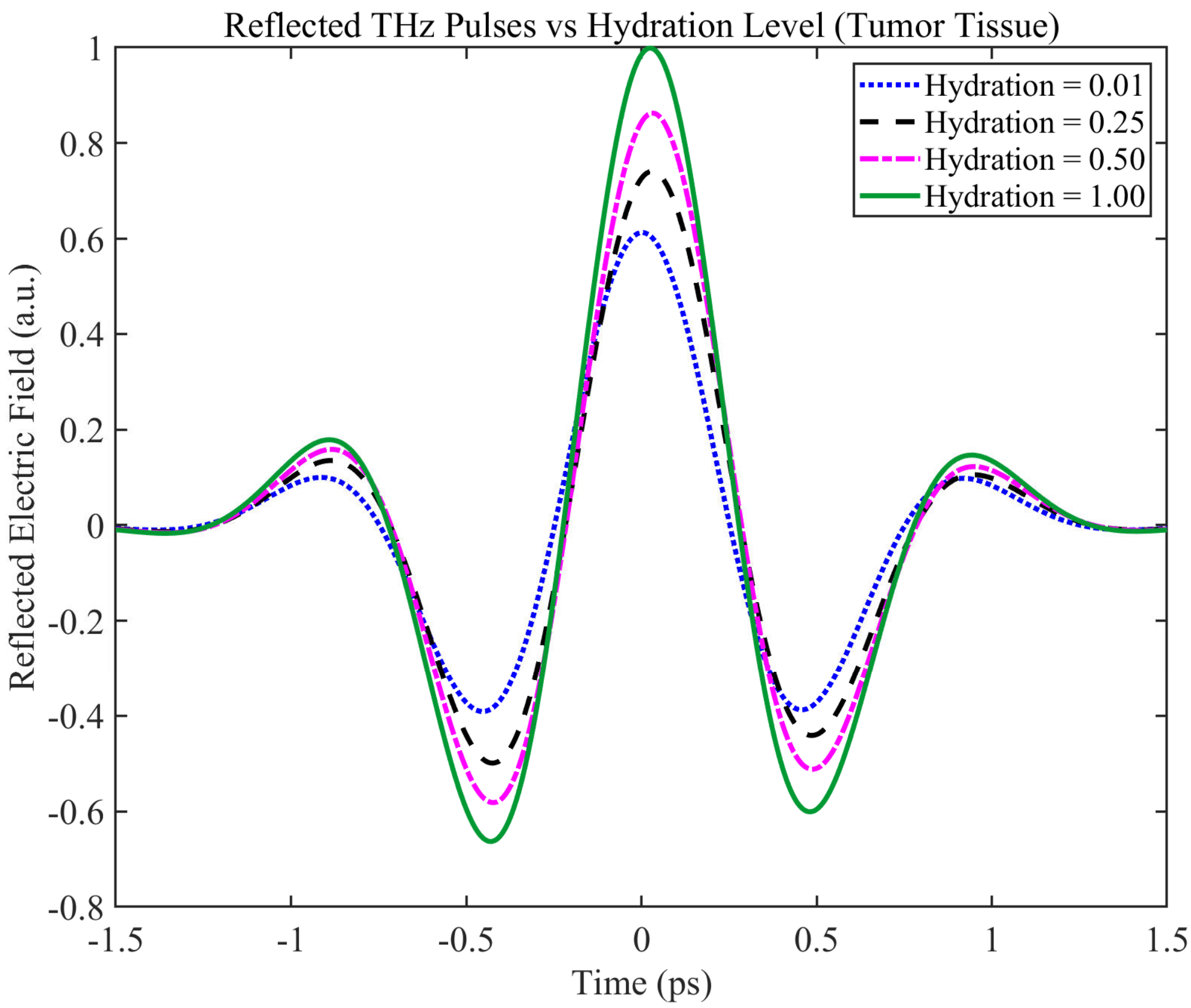

Figure 7: Sensitivity of reflected THz pulse waveforms to hydration level in tumour tissue models.

The Fresnel coefficients that were calculated from the triple-Debye-derived complex refractive indices at different frequencies and angles of incidence is depicted in Figure 8. For all tissues, the reflection coefficient magnitude $|\Gamma|$ decreases steadily with frequency at normal incidence. This is in line with the fact that the refractive index dispersion decreases at greater THz frequencies. Tumor tissue has the highest $|\Gamma|$ for both TE and TM polarizations (about 0.8–0.9 at low frequencies), followed by normal tissue, water, and fat, which have the lowest reflectivity (about 0.3). The close match between the tumor and water curves shows how much greater hydration levels affect malignant areas. The lower reflectivity of healthy tissue that is mostly fat gives a clear starting point for contrast. Notably, TM polarization gives slightly lower $|\Gamma|$ than TE across the band. This is because the impedance definitions are different ($\eta_{TM}$ involves $n$ versus $1/n$ for TE). Panels (c) and (d) show that the angular dependence is more complicated. For TE polarization (panel c), transmission $|T|$ stays pretty high (>1.5) at grazing angles but drops quickly near 90° when total internal reflection starts. Hydrated tissues (tumor and water) have wider and greater peaks because they have larger refractive indices. On the other hand, TM polarization (panel d) shows the Pseudo-Brewster-angle minimum near 60–70°, where $|T|$ approaches unity and $|\Gamma|$ goes away for p-polarized light. This suppression is strongest for water and tumors, and it moves to greater angles as n increases and disappears completely for low-index adipose. These polarization-specific features, especially the Pseudo-Brewster dip and its position that depends on how much water is in it, give you more options for improving tumor contrast in oblique-incidence imaging geometries, which is what handheld THz probes usually use.

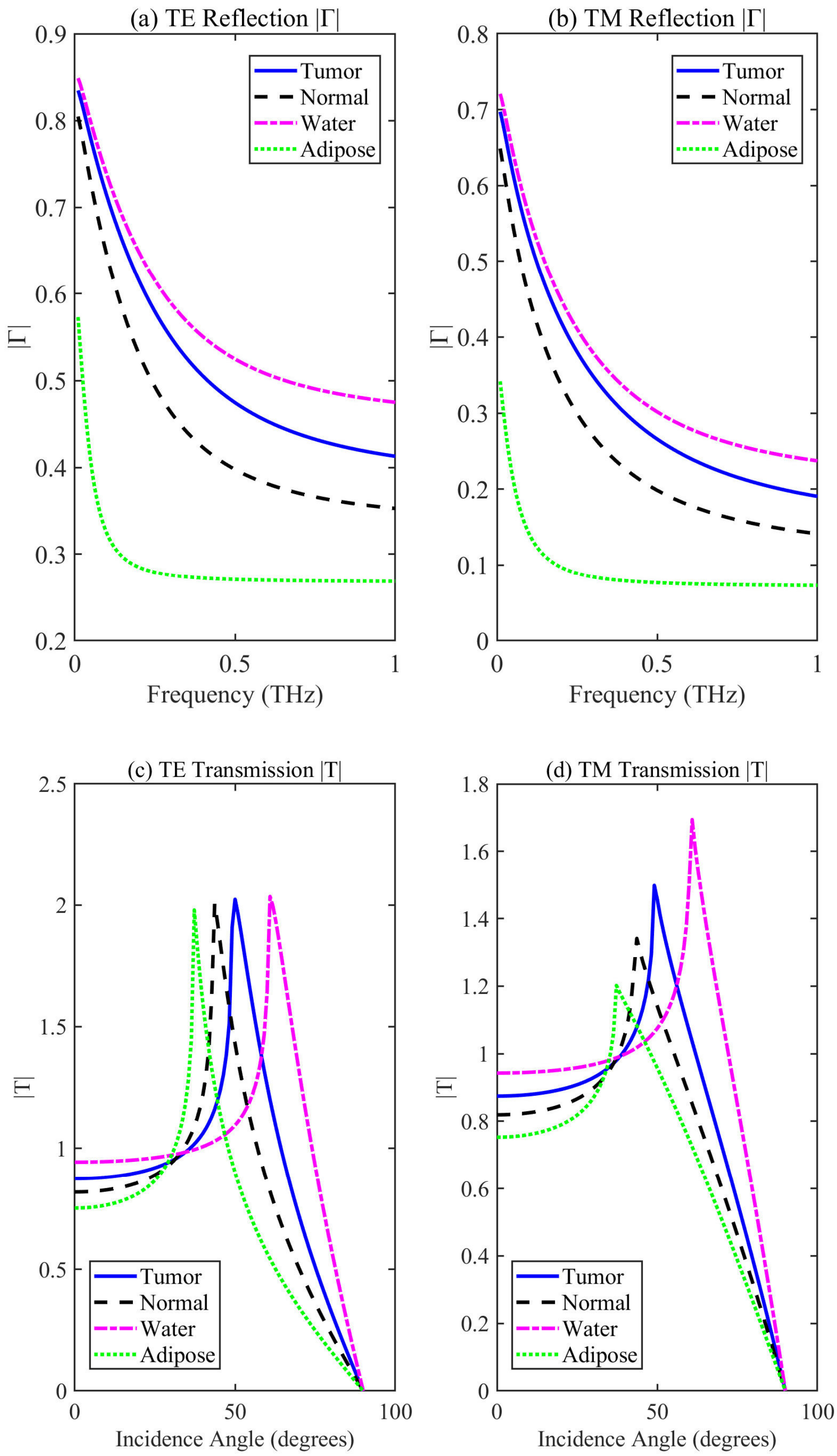

Figure 8: Polarization- and angle-dependent Fresnel reflection and transmission coefficients for different tissue.

Figures 9(a) and 9(b) present two-dimensional frequency–angle maps of the magnitude of the Fresnel reflection coefficient for TE and TM polarizations, respectively, calculated using the fitted triple-Debye dielectric model. The maps simultaneously illustrate the influence of dielectric dispersion, tissue composition, polarization, and incidence angle on THz reflection over the investigated spectral range. For TE polarization, the reflection coefficient varies smoothly with both frequency and incidence angle. All hydrated media exhibit strong reflection at grazing incidence, where the impedance mismatch approaches its maximum and the reflection coefficient approaches unity. As the frequency increases, the gradual reduction of the real permittivity caused by dielectric relaxation decreases the impedance contrast, producing a continuous reduction in reflection magnitude. Tumour tissue consistently maintains a greater reflection level than normal tissue because its larger water content leads to greater permittivity over the entire frequency band. Pure water exhibits a very similar behaviour owing to its dominant free-water relaxation, whereas adipose tissue, characterized by significantly lower dielectric polarization, presents the weakest overall reflection response. The TM polarization maps exhibit a markedly different behaviour due to the polarization dependence of the Fresnel equations. Instead of the smooth variation observed for TE polarization, pronounced low-reflection regions appear as frequency-dependent bands associated with the Pseudo-Brewster condition. Because the dielectric constant decreases with frequency, the Pseudo-Brewster angle also changes continuously, producing the characteristic curved suppression bands visible in the TM maps. These minima are particularly pronounced for adipose tissue owing to its relatively low refractive index, whereas tumour tissue and water retain high reflection over most of the parameter space because their larger dielectric constants shift the Pseudo-Brewster condition toward larger incidence angles and reduce its overall influence. The contrast between tumour and normal tissue remains evident in both polarizations but is governed by different physical mechanisms. Under TE polarization, the contrast primarily originates from the larger impedance mismatch associated with the increased water content of malignant tissue, producing consistently stronger reflected fields. Under TM polarization, the frequency-dependent Pseudo-Brewster suppression introduces an additional polarization-selective contrast mechanism, allowing healthy low-permittivity tissues to exhibit substantially reduced reflection while tumour tissue maintains relatively high reflection. This behaviour suggests that combining TE and TM measurements can enhance tissue discrimination by simultaneously exploiting dielectric contrast and polarization-dependent interface effects.

Figure 10 presents two-dimensional maps of the delay dispersion ($d\tau/df$) as a function of frequency and tissue thickness for tumour tissue, normal breast tissue, pure water, and adipose tissue. The delay dispersion is calculated from the frequency derivative of the propagation delay obtained using the optimized Triple-Debye dielectric model and provides a direct measure of the temporal broadening experienced by broadband THz pulses during propagation through dispersive biological media. For all tissues, the magnitude of the delay dispersion increases with increasing propagation distance, reflecting the cumulative nature of dispersive phase delay. As the tissue thickness increases from sub-millimeter values to 5 mm, progressively larger portions of the broadband THz spectrum experience different propagation velocities, resulting in stronger temporal spreading of the reflected pulse. This behaviour represents a fundamental characteristic of dispersive wave propagation and directly determines the temporal resolution achievable in THz time-domain imaging. A distinct dependence on tissue composition is also observed. Tumour tissue exhibits a relatively moderate variation of delay dispersion across the investigated frequency range, producing smooth spatial gradients that arise from its broadband dielectric relaxation.

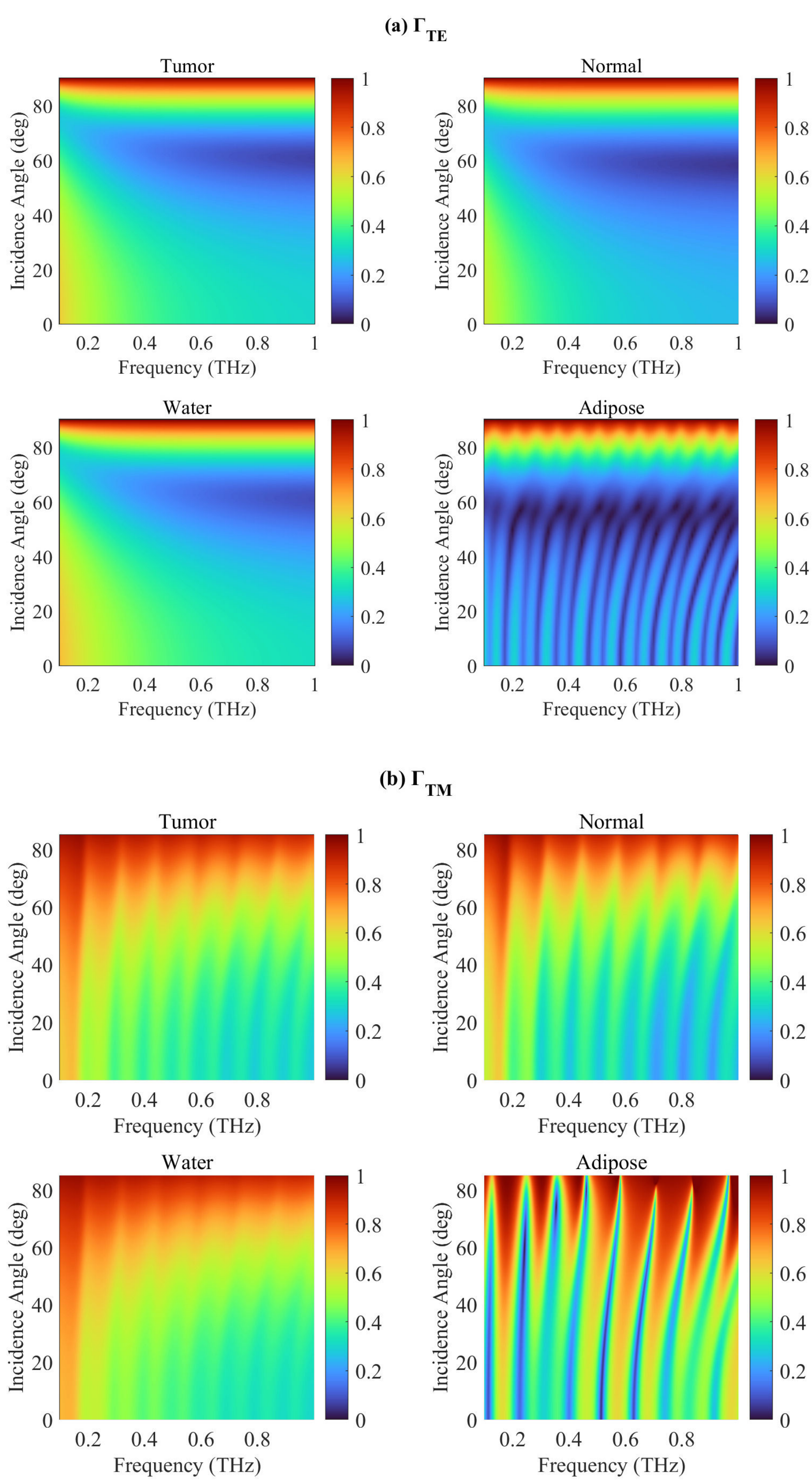

Figure 9: Frequency–angle maps of the magnitude of the Fresnel reflection coefficient for (a) TE and (b) TM polarizations calculated using the triple-Debye dielectric model.

This behaviour reflects the combined contribution of free-water, bound-water, and interfacial polarization processes incorporated in the Triple-Debye model. In comparison, normal breast tissue indicates a substantially larger dispersion magnitude, particularly at low frequencies and greater propagation depths, indicating stronger frequency-dependent group delay accumulation within the investigated parameter range. Pure water displays a distinct behaviour characterized by a continuous increase in dispersion toward greater frequencies. Because water possesses the strongest molecular relaxation among the investigated materials, the frequency dependence of the group velocity becomes increasingly pronounced as propagation distance increases, producing the largest high-frequency temporal spreading. In contrast, adipose tissue exhibits comparatively weak dielectric dispersion over most of the spectrum owing to its low water content and reduced polarization strength. Therefore, the delay dispersion rapidly approaches nearly constant values at greater frequencies, indicating that THz pulses experience relatively limited temporal distortion during propagation through adipose-rich regions. An significant feature visible in all four maps is the gradual transition from strong low-frequency dispersion toward a more saturated behaviour above approximately 2–3 THz. This trend reflects the progressive reduction of dielectric relaxation as the material approaches its high-frequency permittivity limit, where variations in the group refractive index become smaller. Additional increases in frequency produce diminishing changes in propagation delay despite continued increases in propagation distance. From an imaging perspective, these results demonstrate that both tissue composition and propagation distance significantly influence the temporal fidelity of reflected THz pulses. Regions exhibiting larger delay dispersion produce stronger pulse broadening and greater waveform distortion, whereas tissues with weaker dispersion preserve the incident pulse shape more effectively. These differences directly affect echo separation, depth resolution, and parameter estimation in THz time-domain reflection systems. The frequency–size maps therefore provide useful guidelines for selecting appropriate operating frequencies and penetration depths that maximize diagnostic contrast while minimizing dispersive degradation of the reflected waveform.

The dominant role of tissue hydration in determining THz attenuation is quantified through parametric maps of the absorption coefficient derived from the proposed triple-Debye model. Figure 11 illustrates the variation of the absorption coefficient as a function of frequency and water fraction for tumour, normal, water, and adipose tissues. A distinct monotonic increase in absorption is observed with increasing water content for all tissue types, confirming that hydration is the primary factor governing dielectric loss in the THz regime. Among all tissues, pure water exhibits the highest absorption coefficients, reaching values of approximately $150cm^{-1}$, reflecting its dominant free-water relaxation response. Tumour tissue shows the second-highest absorption, with peak values approaching $100cm^{-1}$, whereas normal fibroglandular tissue reaches approximately $60cm^{-1}$. Adipose tissue exhibits the weakest absorption, remaining below $15cm^{-1}$ over the entire parameter range because of its low water content and lipid-rich composition. For all tissue types, absorption increases with increasing water fraction while exhibiting a gradual frequency dependence governed by the dielectric relaxation processes incorporated in the triple-Debye model. The tumour panel displays a noticeably steeper increase with hydration than normal tissue, indicating that relatively small increases in tissue water content produce measurable increases in THz attenuation. This behaviour is consistent with the elevated intracellular and extracellular water content typically observed in malignant breast tissue. These absorption characteristics are directly linked to the time-domain responses presented in the previous figures. Increased dielectric loss within tumour tissue reduces the transmitted field while simultaneously enhancing the dielectric contrast responsible for stronger reflected waveforms.

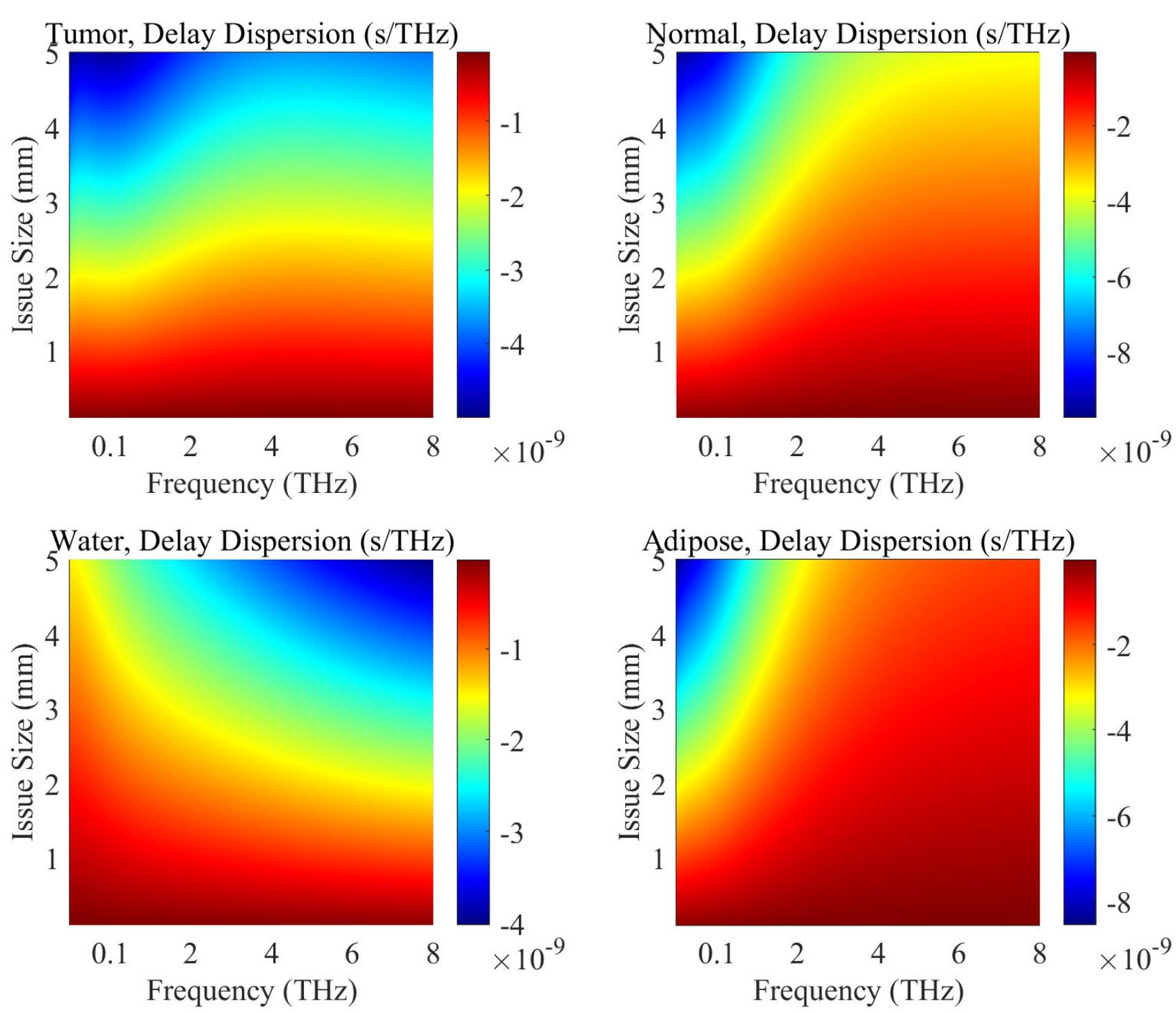

Figure 10: Frequency–size maps of the delay dispersion $(d\tau/df)$ predicted by the optimized Triple-Debye dielectric model for tumour tissue, normal breast tissue, pure water, and adipose tissue.

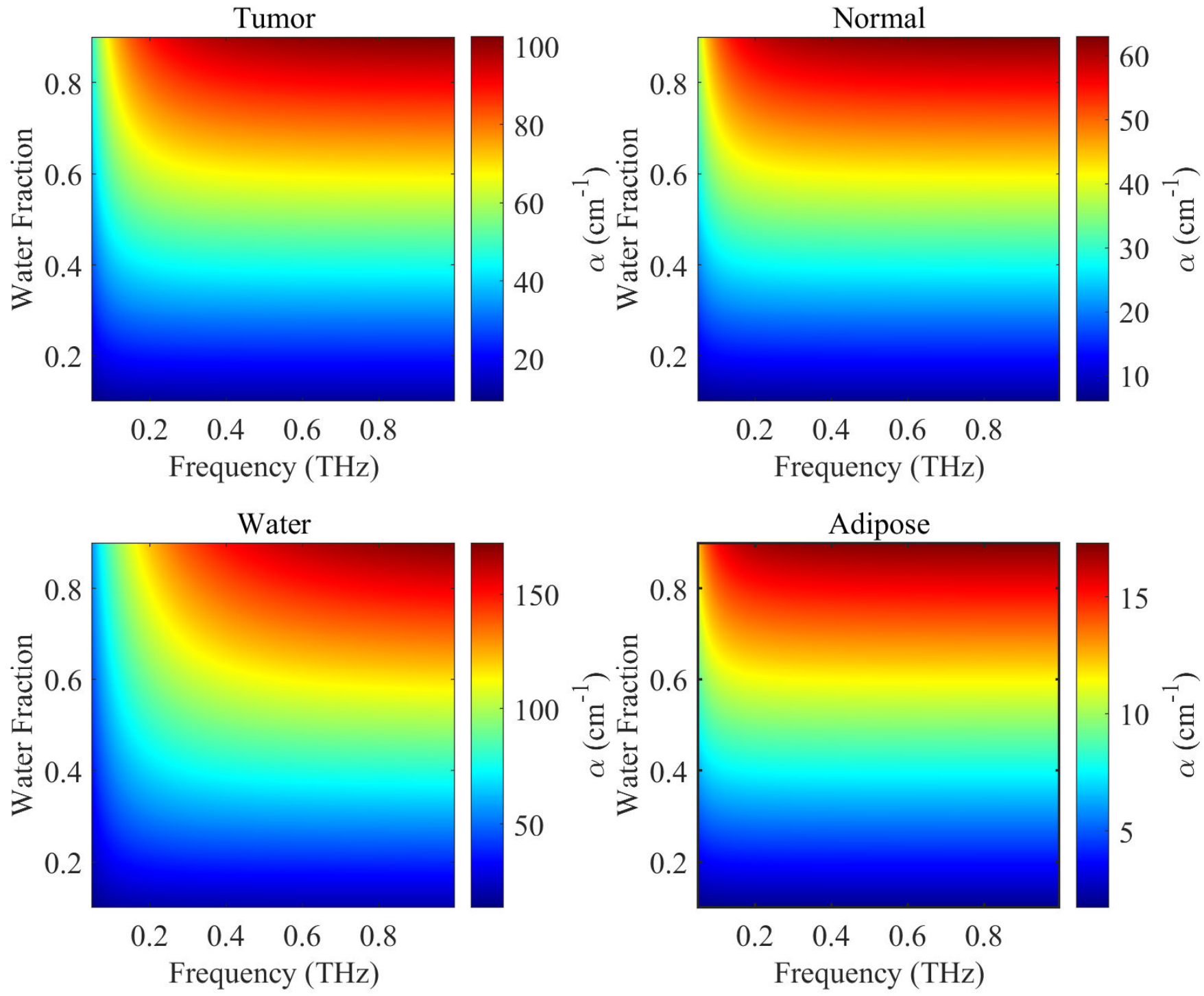

Figure 11: Heatmaps of the frequency-dependent absorption coefficient as a function of water fraction for various tissues derived from triple-Debye parameters.

The practical implications of THz–tissue interactions for electromagnetic energy deposition are illustrated by the absorbed power fraction within the tissue layer. Figure 12 presents two-dimensional maps of the absorbed power fraction for tumour and normal breast tissue under both TE and TM polarizations as functions of frequency and incidence angle. Under TE polarization, tumour tissue exhibits substantially greater absorption than normal tissue over a broad range of frequencies and incidence angles, particularly at low frequencies and moderate incidence angles. This behavior originates from the larger dielectric loss of hydrated malignant tissue, reflected by its greater imaginary permittivity (Figure 11), which converts a greater portion of the incident electromagnetic energy into dielectric dissipation. As the frequency and incidence angle increase, the absorbed power gradually decreases because the interaction time within the tissue becomes shorter and the effective coupling into lossy relaxation mechanisms is reduced. In contrast, normal breast tissue exhibits consistently lower absorbed power owing to its reduced water content and weaker dielectric relaxation. Under TM polarization, the absorbed power is generally lower for both tissue types than under TE polarization. Broad low-absorption regions appear around incidence angles of approximately $50° - 70°$, corresponding to the hydration-dependent pseudo-Brewster regime where reflection is significantly reduced and a larger fraction of the incident field is transmitted through the tissue interface. Consequently, less electromagnetic energy is dissipated within the tissue despite the increased field penetration. These absorption maps provide additional physical insight into the polarization dependence of THz propagation in heterogeneous breast tissue. TE polarization enhances dielectric energy deposition within highly hydrated malignant tissue while simultaneously producing stronger reflected signals, whereas TM polarization favors energy transmission through the interface and therefore reduces both reflection and absorption around the pseudo-Brewster condition. The proposed triple-Debye dielectric framework successfully reproduces these coupled polarization-dependent interactions between dielectric loss, reflection, transmission, and energy deposition over a broad THz frequency range. These findings provide a physically consistent basis for the design and future experimental evaluation of polarization-sensitive THz reflection imaging systems for breast tissue characterization.

The limited penetration of THz radiation in hydrated tissues, while a challenge for deep imaging, constitutes a key advantage for surface-sensitive applications such as tumour margin assessment. In this respect, figure 13 quantifies penetration depth and the impact of overlying adipose layers on detectable signals. The heatmaps show that the penetration depths in tumor tissue are less than 0.17 mm for most frequencies and angles. This is because the water content is consistently high, which means that there is little variation. Normal breast tissue allows for deeper penetration (approximately 0.15–0.3 mm), especially at elevated frequencies where absorption diminishes, indicating reduced average hydration in mixed fibroglandular-adipose composition. This sub-millimeter confinement guarantees that reflected THz pulses primarily sample the immediate tissue surface, rendering the technique inherently effective. The lower panel shows how the thickness of the fat layer (2–8 mm) affects the reflected pulse amplitude over time. This is because of the adipose tissue that is present in the breast. Even a 2 mm adipose overlay significantly reduces the primary reflection from the tissue below it. The amplitude drops sharply for thicker layers (5 mm and 8 mm). This decay, which looks like an exponential curve, happens due to cumulative absorption and scattering in the low-loss but dispersive fat medium. This effectively suppresses late-time echoes from more distant interfaces. These results indicate that the high refractive index of hydrated malignant tissue produces enhanced surface reflectivity, whereas its stronger dielectric absorption simultaneously limits electromagnetic penetration depth.

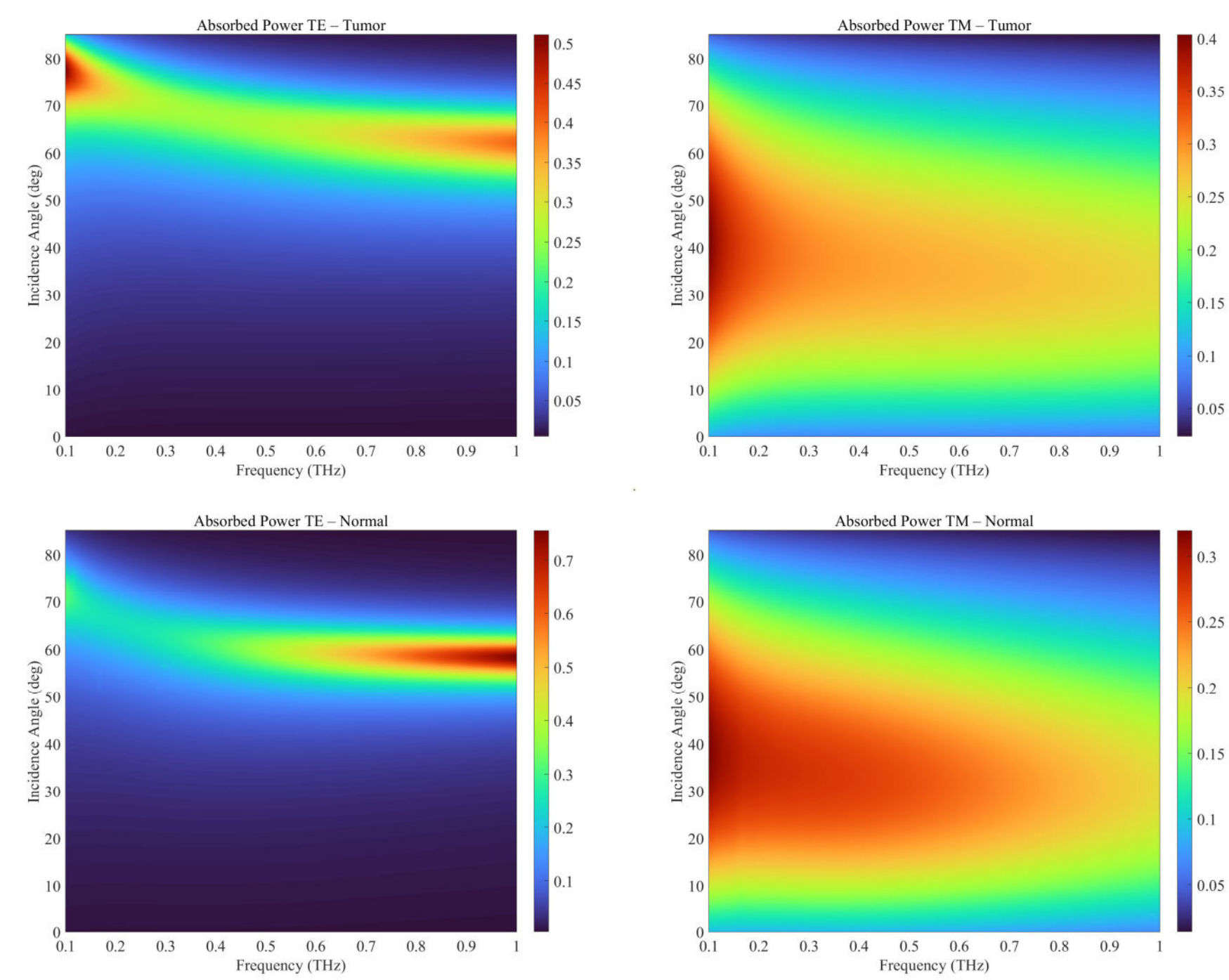

Figure 12: Heatmaps of absorbed power fraction in tumour and normal breast tissue for TE and TM polarizations as a function of frequency and incidence angle.

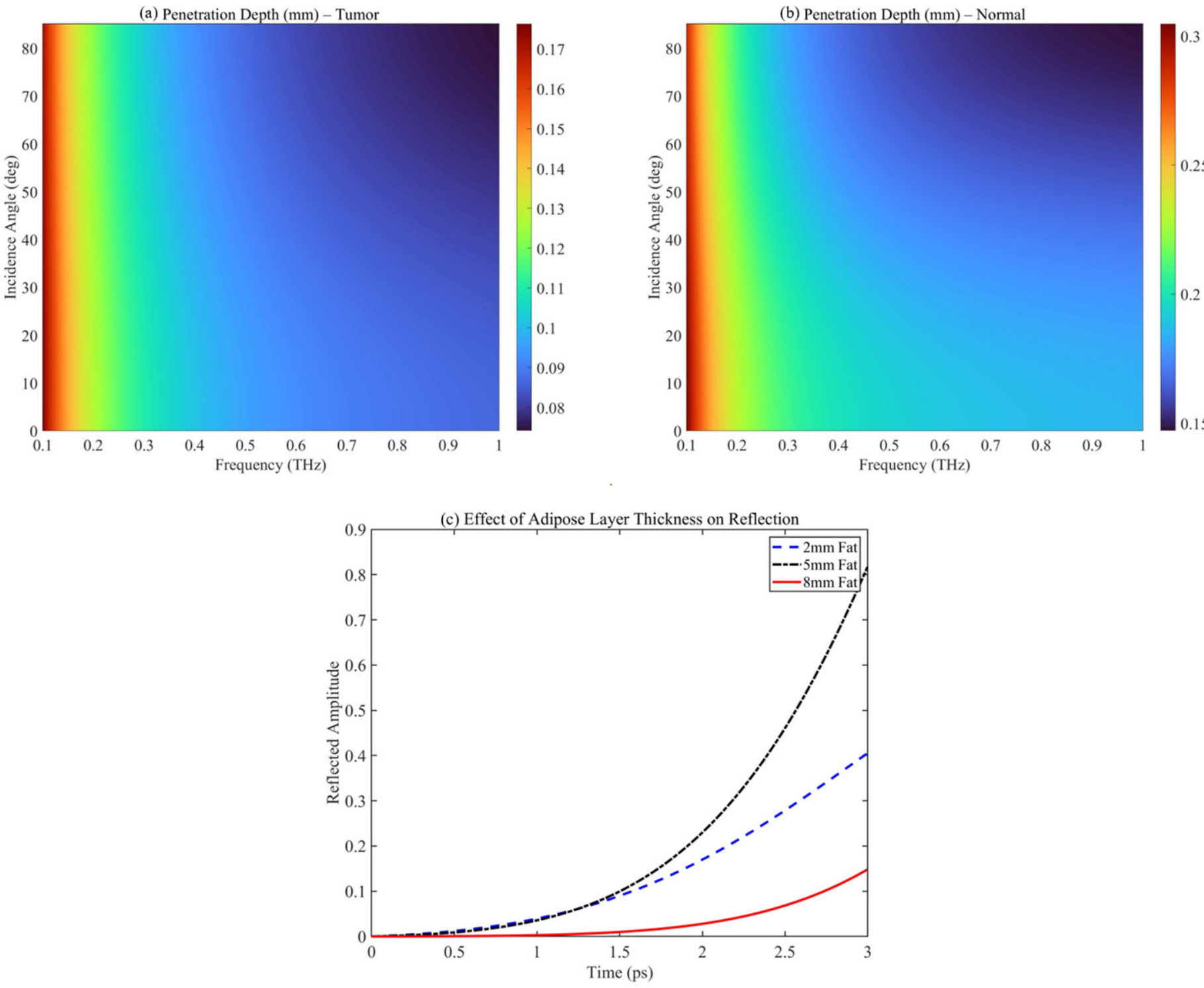

Figure 13: (a, b) Heatmaps of THz penetration depth(mm) in tumour and normal breast tissue versus frequency and incidence angle, (c) the effect of overlying adipose layer thickness on reflected pulse amplitude over time.

Figure 14 presents two-dimensional maps of the predicted signal-to-noise ratio (SNR) for tumour and normal breast tissues as functions of operating frequency and acquisition (integration) time. The calculations combine the reflected field predicted by the proposed triple-Debye dielectric model with a detector noise model in which the noise level decreases proportionally to $1/\sqrt{T_{int}}$, thereby accounting for the improvement in measurement quality achieved through temporal averaging. For both tissue types, the SNR increases monotonically with increasing integration time because longer acquisition intervals reduce the influence of random detector noise. A similar improvement is observed with increasing frequency throughout most of the investigated spectral range. The strongest frequency dependence occurs below approximately 0.4 THz, where relatively small increases in frequency produce a substantial enhancement in SNR. Above approximately 0.6 THz, however, the colour gradients become progressively weaker, indicating that further increases in frequency provide only marginal improvements in measurement quality. The tumour map consistently exhibits greater SNR values than the normal tissue map over the entire frequency and integration-time domain. This behaviour originates from the stronger dielectric contrast and larger Fresnel reflection coefficient predicted for malignant tissue by the triple-Debye model. Although tumour tissue also experiences greater dielectric absorption because of its greater water content, the increase in interface reflectivity dominates within the investigated frequency range, resulting in a stronger received reflected signal and consequently a greater SNR under identical acquisition conditions. The contour distributions also reveal that the dependence on integration time gradually weakens as the acquisition duration increases. Beyond approximately 5–6 s, the colour variation along the vertical direction becomes relatively small, suggesting that the SNR approaches a saturation regime in which additional averaging yields diminishing returns. This behaviour is consistent with the expected square-root dependence of SNR on integration time and indicates that moderate acquisition durations are sufficient to obtain stable measurements for both tissue types. From a practical perspective, these results suggest that operating within the intermediate frequency range of approximately 0.4–0.8 THz provides an effective compromise between measurement sensitivity and acquisition efficiency. Within this region, tumour tissue exhibits consistently greater SNR than normal tissue while avoiding the limited improvements observed at the highest frequencies. Consequently, the proposed triple-Debye framework not only reproduces the experimentally observed dielectric behaviour of breast tissue but also enables quantitative prediction of system-level imaging performance, providing useful guidance for optimizing operating frequency and acquisition time in reflection-mode THz breast imaging systems.

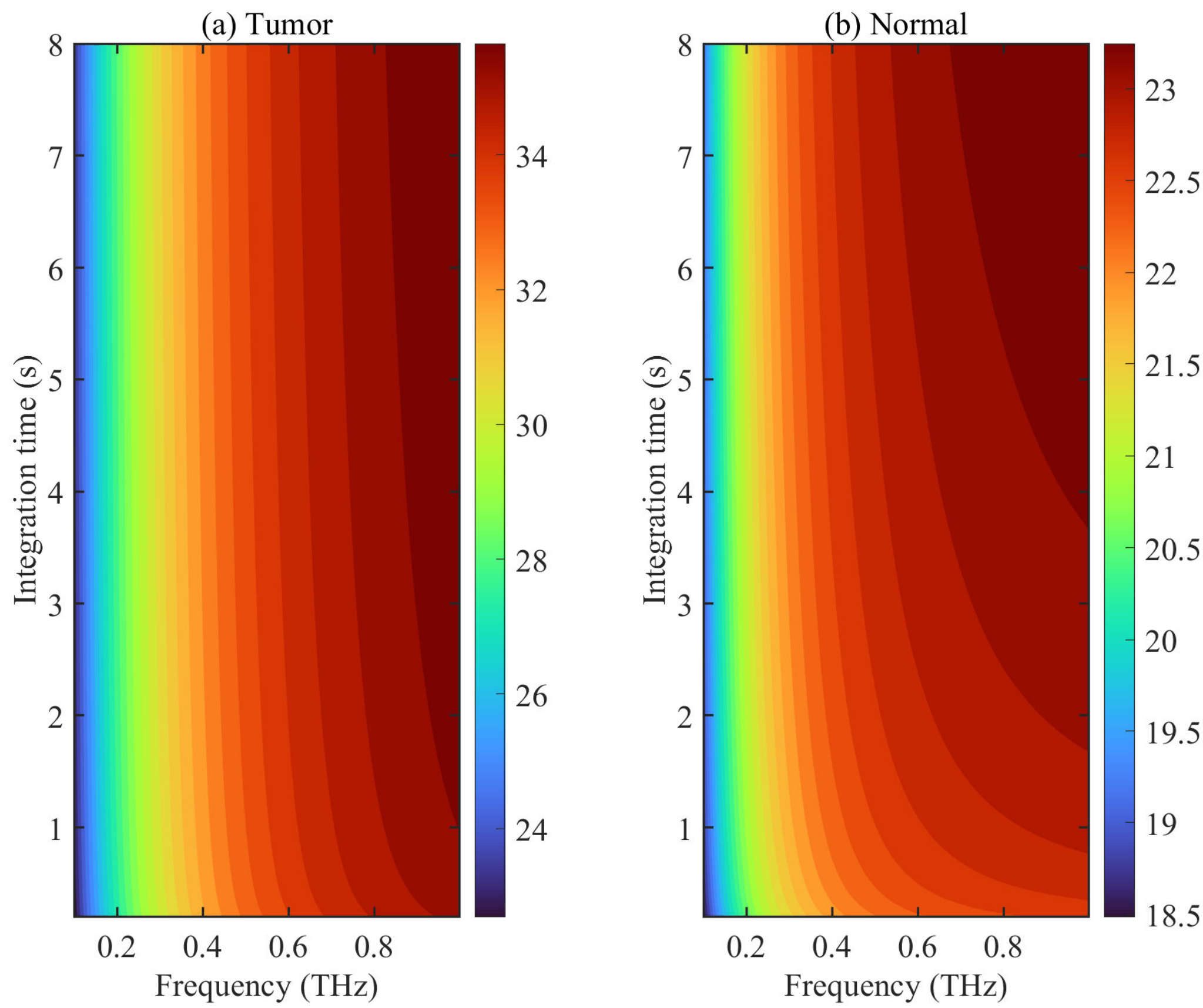

Figure 14: Heatmaps of Predicted signal-to-noise ratio (SNR) maps for reflection-mode THz imaging of (a) tumour and (b) normal breast tissue as functions of operating frequency and acquisition (integration) time.

# 4 Conclusion

This work establishes a physically motivated, multi-scale dielectric framework for quantitatively predicting broadband THz interactions with heterogeneous breast tissue. By explicitly incorporating three physically distinct relaxation processes associated with free-water rotational dynamics, bound-water relaxation, and ultrafast interfacial/macromolecular polarization, the proposed triple-Debye model provides an accurate representation of the multi-step dielectric dispersion and broadband absorption observed in *ex vivo* human breast tissue over a broad THz frequency range. Nonlinear least-squares fitting of experimentally measured refractive-index data demonstrates a substantial improvement over conventional dielectric models, reducing the fitting root-mean-square error (RMSE) from 0.7773 (single Debye) and 0.2976 (double Debye) to only 0.0199 for the proposed triple-Debye formulation. These results confirm that inclusion of the third relaxation process is essential for accurately describing the intermediate-frequency dielectric response associated with tissue hydration. Full-wave FDTD-ADE simulations of realistic multilayer breast structures further demonstrate that reflected THz waveforms encode tissue hydration through distinct and reproducible temporal characteristics, including greater reflection amplitudes, delayed pulse arrival, increased waveform broadening, and enhanced late-time oscillations in malignant tissue relative to normal fibroglandular tissue. These waveform differences originate from the combined effects of increased refractive-index contrast and stronger frequency-dependent dielectric relaxation associated with elevated water content in malignant tissue, confirming hydration as the dominant physical mechanism governing THz contrast. Polarization- and angle-resolved Fresnel analysis further clarifies the underlying contrast mechanisms. While TE polarization maintains relatively strong reflection over a broad angular range, TM polarization near the hydration-dependent pseudo-Brewster minimum selectively suppresses reflections from low-hydration normal and adipose tissue, providing an additional physics-based mechanism for polarization-dependent contrast. Furthermore, the proposed framework accurately predicts frequency-dependent attenuation, penetration depth, interface reflectivity, and signal-to-noise characteristics. The simulations indicate that overlying adipose layers act as natural low-pass filters by attenuating late-time echoes originating from deeper interfaces, whereas the predicted signal-to-noise ratios suggest that acquisition times on the order of several tens of seconds are achievable under the simulated system assumptions, providing useful guidance for the design and optimization of future experimental THz reflection imaging platforms. This proposed triple-Debye framework reproduces broadband dielectric behavior considerably more accurately than lower-order Debye models while providing physically interpretable electromagnetic signatures associated with tissue hydration. By integrating experimentally validated dielectric modeling with full-wave electromagnetic simulations, this study provides a computational framework for future experimental investigations of polarization-sensitive THz reflection imaging.

**Acknowledgment & Funding**

This work did not receive any specific grant from funding agencies in the public, commercial, or not-for-profit sectors.

The authors acknowledge Ms. Farkhonde Zamaninejad for her contribution during the preliminary stage of this work and the preparation of the initial conference abstract.

**Author Contributions**

Ali Asghar Molavi Choobini was responsible for conceptualization, data curation, Software, formal analysis, investigation, methodology, writing the original draft, review and editing.

M. Shahmansouri contributed equally to investigation, project administration, supervision, validation, and review and editing of the manuscript

**Data availability statement**

The main results and processed data are presented in the published article. Additional simulation outputs and supporting materials are available from the corresponding author upon reasonable request.

**Competing interests**

The authors declare no competing interests.